\documentclass[aps,pra,superscriptaddress,showpacs,floatfix,twocolumn]{revtex4-2}
\usepackage[dvips]{graphicx}

\usepackage{longtable}
\usepackage{dcolumn}
\usepackage[dvips]{graphicx}
\usepackage{bm}
\usepackage{lipsum}
\usepackage{bbm}
\usepackage{color}

\usepackage{times}
\usepackage{nicefrac}
\usepackage{amsmath}
\usepackage{amsfonts}
\usepackage{amssymb}
\usepackage{amsthm}
\usepackage[normalem]{ulem}
\newcolumntype{.}{D{x}{}{-1}}

\def\ketm#1{  \left\vert  #1   \right\rangle   }

\def\mem#1#2#3{  \left\langle #1 \left\vert  #2 \right\vert #3 \right\rangle   }

\begin{document}

\newcommand{\vare}{\varepsilon}

\newcommand{\pr}{^{\prime}}
\newcommand{\bfa}{{\bf a}}
\newcommand{\bfx}{{\bf x}}
\newcommand{\bfe}{{\bf e}}
\newcommand{\bfy}{{\bf y}}
\newcommand{\bfr}{{\bf r}}
\newcommand{\bfR}{{\bf R}}
\newcommand{\bfz}{{\bf z}}
\newcommand{\bfp}{{\bf p}}
\newcommand{\la}{\langle}
\newcommand{\ra}{\rangle}
\newcommand{\eps}{\varepsilon}
\newcommand{\beq}{\begin{equation}}
\newcommand{\eeq }{\end{equation}}
\newcommand{\beqn}{\begin{eqnarray}}
\newcommand{\eeqn }{\end{eqnarray}}
\newcommand{\ba}{\begin{array}}
\newcommand{\ea}{\end{array}}
\newcommand{\balpha}{{\mbox{\boldmath$\alpha$}}}
\newcommand{\az}{\alpha Z}
\newcommand{\aZ}{\alpha Z}
\newcommand{\etal}{{\it et al. }}

\newcommand{\lbr}{\langle}
\newcommand{\rbr}{\rangle}

\newcommand{\Dmatrix}[4]{
        \left(
        \begin{array}{cc}
        #1  & #2   \\
        #3  & #4   \\
        \end{array}
        \right)
        }

\title{{\it Ab initio} QED calculations in diatomic quasimolecules}

\author{A.~N.~Artemyev}
\affiliation{Institut f\"{u}r Physik und CINSaT, Universit\"{a}t Kassel,
  Heinrich-Plett-Stra{\ss}e 40, 34132 Kassel, Germany}

\author{A.~Surzhykov}
\affiliation{Physikalisch--Technische Bundesanstalt, D--38116 Braunschweig, Germany}
\affiliation{Institut f\"ur Mathematische Physik, Technische Universit\"at Braunschweig, D--38106 Braunschweig, Germany}
\affiliation{Laboratory for Emerging Nanometrology Braunschweig, D-38106 Braunschweig, Germany}

\author{V.~A.~Yerokhin}
\affiliation{Physikalisch--Technische Bundesanstalt, D--38116 Braunschweig, Germany}

\begin{abstract}
We present a theoretical approach for {\it ab initio} calculations of the one--loop QED corrections to energy levels of heavy diatomic quasimolecules. This approach is based on the partial--wave expansion of the molecular wave and Green functions in the basis of monopole solutions, written in spherical coordinates. By using so generated molecular functions we employed the existing atomic--physics techniques to evaluate the self--energy and vacuum--polarization corrections. In order to illustrate the application of our method, we perform detailed calculations of the Dirac energy and QED corrections for the 1$\sigma_g$ ground state of homonuclear U$_2^{183+}$ as well as heteronuclear U--Pb$^{173+}$ and Bi--Au$^{161+}$ quasimolecules. 
\end{abstract}
\maketitle
\section{Introduction}

High--precision calculations of energy levels of atoms and ions are impossible today without a proper account for the quantum electrodynamics (QED) corrections. The theory, that describes QED effects for an electron \textit{bound} in the Coulomb field of a nucleus, has been extensively studied over decades \cite{mohr:98, yerokhin:15}. With the help of the bound--state QED approach, detailed calculations have been performed both for light atomic systems and also for medium-- and even high--$Z$ ions \cite{kaygorodov:19,kozhedub:19,yerokhin:20}. The results of these calculations were found to be in a good agreement with experimental data and provided valuable insight into the physics of strong electromagnetic fields \cite{Indelicato:19}.

In contrast to atoms, less progress has been made so far in developing an efficient QED theory to describe energies of bound states of molecules and molecular ions. Accurate calculations of QED effects were performed for the lightest diatomic molecules (H$_2$, HD, etc.) and molecular ions (H$_2^+$, HD$^+$, etc.), see Refs.~\cite{zhong:09,beyer:19,komasa:11,korobov:14,puchalski:18,puchalski:19,karr:20,korobov:21}. These calculations were carried out within the approach based on the expansion in the parameter $\alpha Z$, where $\alpha$ is the fine-structure constant and $Z$ is the nuclear charge number. The region of applicability of this approach is restricted to light systems, for which $\alpha Z << 1$ is a small parameter. During the recent years, however, considerable interest has arisen to explore QED effects in medium-- and high--$Z$ molecular systems. In particular, a further increase of accuracy of quantum chemistry calculations for heavy molecules requires the consideration of QED corrections \cite{pyykko:12,HRQC:17}.

A deep theoretical understanding of QED effects is also highly demanded for investigations of quasimolecules, i.e. short--lived dimers that are formed in slow collisions of highly--charged heavy ions with atomic (or ionic) targets. The quasimolecules are considered today as a unique tool to explore instability of the QED vacuum in extremely strong electromagnetic fields, produced by colliding nuclei \cite{pieper:69,greiner:85}. In the past, a series of experiments have been performed at the GSI facility in Darmstadt to observe the formation of quasimolecules in Bi$^{q+}$--Au and U$^{q+}$--Au (ion--atom) collisions \cite{verma:05,verma:06:rphys,verma:06:nimb}. Even more advanced studies, including ion--ion U$^{91+}$--U$^{92+}$ encounters, are planned at the Facility for Antiproton and Ion Research (FAIR). The guidance and analysis of these experiments will require high--precision calculations of quasimolecular energy levels and, hence, accounting for the QED corrections.

It is a challenging task to evaluate QED corrections for (quasi) molecular systems, that consist of two and even more Coulomb centers and, hence, do not posses spherical symmetry. In our previous work \cite{artemyev:15} we dealt with this problem and proposed a two--step \textit{ab initio} approach in which (i) molecular wave functions are generated first for the monopole (spherically--symmetric) case, and (ii) later used as a basis to construct eigensolutions of the two--center Dirac Hamiltonian. Being developed in spherical coordinates, this approach allows one to use the well--elaborated atomic--physics techniques to calculate QED corrections to molecular energy levels to all orders in $\alpha Z$. In order to illustrate the application of the proposed theory, we have computed the QED corrections to the energy of the ground state of U$^{91+}$--U$^{92+}$ (also denoted as U$_2^{183+}$) quasimolecule \cite{artemyev:15}. Until now, however, the theoretical analysis has been restricted to this particular homonuclear case only. In the present work, we extend our approach to explore heteronuclear quasimolecules, such as U--Pb$^{173+}$ or Bi--Au$^{161+}$, that are of particular interest for experimental investigations of super--critical phenomena. Moreover, we provide the detailed derivation of the formulas omitted in Ref.~\cite{artemyev:15} and introduce new criteria to estimate the accuracy of our predictions.

The paper is organized as follows. In the Section~\ref{subsec:wave_functions} we briefly recall the approach to construct (two--center) molecular wave and Green functions in terms of their monopole counterparts. These functions, written in spherical coordinates, are used in Section~\ref{subsec:QED_corrections} to evaluate the first--order self--energy and vacuum--polarization QED corrections to the energy levels. The details of numerical algorithms and of uncertainty estimations, as well as the results of our calculations are presented then in Section~\ref{result}. Here, in particular, we report results for the zero--order (Dirac) energy and QED corrections for the 1$\sigma_g$ ground state of U$_2^{183+}$, U--Pb$^{173+}$ and Bi--Au$^{161+}$ quasimolecules. Results of our calculations indicate that the proposed approach allows a computation of QED corrections with an accuracy varying from 0.1~\% for small inter--nuclear distances to about 10~\% for cases when nuclei are displaced very far from each other. The summary of these results and a short outlook are given finally in Section~\ref{summary}.

The relativistic units ($\hbar = c = m_e = 1$) are used throughout the paper if not stated otherwise.

\section{Theoretical background}
\label{theory}

\subsection{Molecular wave and Green's functions}
\label{subsec:wave_functions}

\begin{figure}[t]
	\includegraphics[width=0.95\linewidth]{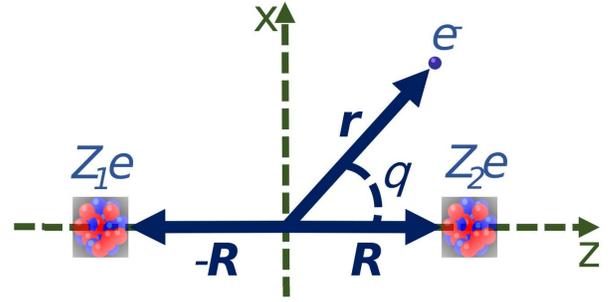}
	\caption{Geometry of the dimer quasimolecule. Positions of the electron and both nuclei are given by the vectors ${\bm r}$, -${\bm R}$ and ${\bm R}$, respectively. The quantization ($z$–-) axis is chosen along the inter–-nuclear axis, and the coordinate origin is taken at the midpoint between the nuclei.}
    \label{Fig1}
\end{figure}

Any theoretical analysis of the QED corrections to the energy levels of diatomic quasimolecules requires the knowledge of the wave functions of an electron, moving in the field of two nuclei. We find these wave functions as solutions of a single--particle Dirac Hamiltonian:
\begin{equation}
    \label{eq:Dirac}
    \hat{H} = \balpha \cdot \bfp + V_{2c}(\bfr) + \beta m_e \, ,
\end{equation}
with the electron--nuclei interaction potential:
\begin{eqnarray}
    \label{eq:total_potential}
    V_{2c}(\bfr) = &-& \frac{\alpha Z_1(1+y_1(|\bfr-\bfR|))}{|\bfr-\bfR|}
    \nonumber \\[0.2cm]
    &-& \frac{\alpha Z_2(1+y_2(|\bfr+\bfR|))}{|\bfr+\bfR|} \, .
\end{eqnarray}
Here we assumed that both nuclei are located on the $z$--axis at distance $R$ from the coordinate origin, chosen at the midpoint between them. The nuclear coordinate vectors, $\pm {\bm R} = \left(0, 0, \pm R\right)$ are directed in this case parallel (antiparallel) to $z$ axis as shown in Fig.~\ref{Fig1}. In Eq.~(\ref{eq:total_potential}), moreover, $y_1$ and $y_2$ are functions, induced by the nuclear charge distribution, which describe the deviation of the nuclear potential from the point--nucleus case.

By using the axial symmetry of the system ``two nuclei \textit{plus} electron'', it is practical to expand the two--center potential (\ref{eq:total_potential}) in terms of Legendre polynomials:
\begin{equation}
    \label{eq:potential_expansion}
    V_{2c}(\bfr) = \sum\limits_{l=0}^\infty V_l(r) \, P_l(\cos \theta) \, ,
\end{equation}
and where the expansion coefficients are:
\begin{equation}
    \label{eq:expansion_coefficients}
    V_l(r) = \frac{2l+1}{2}\int\limits_0^\pi \sin \theta\, d\theta\ V_{2c}(r, \cos \theta)
P_l(\cos \theta) \, .
\end{equation}
Here $\theta$ is the polar angle of the electron with respect to the internuclear axis, see Fig.~\ref{Fig1}, and $V_{2c}(r, \cos \theta) \equiv V_{2c}(\bfr)$.

As seen from Eq.~(\ref{eq:potential_expansion}), the two--center potential $V_{2c}(\bfr)$ can be presented as a sum of (i) the spherically--symmetric monopole term with $l=0$, and (ii) the higher multipole contributions, that depend on $\theta$. This decomposition of $V_{2c}(\bfr)$ allows one to apply the two--step procedure to generate eigenfunctions of the Hamiltonian (\ref{eq:Dirac}). Since the details of this procedure have been presented in our previous work \cite{artemyev:15}, here we just briefly recall the basic ideas. At the first step we use the dual--kinetically balanced B--spline approach \cite{shabaev:04:dkb,johnson:88} to  solve the Dirac equation in the monopole approximation, i.e. for $V_{2c}(r, \cos \theta) = V_{l = 0}(r)$. The quasi--complete set of eigenenergies and eigenfunctions, $\{\epsilon_{n,\kappa,\mu}\}$ and $\{\phi_{n,\kappa,\mu}(\bfr)\}$, obtained from the B--spline approach, are characterized by the principal and Dirac quantum numbers $n$ and $\kappa$, as well as by the projection $\mu$ of the total angular momentum of electron $j = |\kappa| - 1/2$ onto the quantization axis. Of course, for the spherically symmetric (monopole) potential the solutions with the same $n$ and $\kappa$ but with different $\mu$'s will have equal radial parts and energies $\epsilon_{n,\kappa} \equiv \epsilon_{n,\kappa,\mu}$.

%
%
%
%

At the second step we expand the eigensolutions $\Phi_{N,\mu}(\bfr)$ of the Dirac Hamiltonian (\ref{eq:Dirac}) with the \textit{full} two--center potential $V_{2c}({\bm r})$ in terms of the monopole solutions:
\begin{equation}
    \label{eq:expansion_PSI}
    \Phi_{N ,\mu}(\bfr) = \sum\limits_{n,\kappa=-K}^K A_{n,\kappa}^{N ,\mu} \, \phi_{n,\kappa,\mu}(\bfr) \, .
\end{equation}
Here, $N$ is just the number of the solution, while $\mu$ is the projection of the total angular momentum that is conserved for the axial symmetry of diatomic (quasi) molecules. In order to accelerate the numerical procedure, the sum in Eq.~(\ref{eq:expansion_PSI}) is restricted to the (monopole) states $\phi_{n,\kappa,\mu}(\bfr)$ with energies $\left| \epsilon_{n,\kappa,\mu} \right| < 100 \; {\rm mc}^2$ and with Dirac quantum number in the range $-K \le \kappa \le K$. The latter restriction provides us also the upper limit $l_{max} = 2K$ for the multipole decomposition of the two--center potential (\ref{eq:potential_expansion}). Finally, the expansion coefficients $A_{n,\kappa}^{N, \mu}$ in Eq.~(\ref{eq:expansion_PSI}) and the eigenenergies ${\cal E}_{N, \mu}$ of the full two--center Hamiltonian (\ref{eq:Dirac}) for each value of $\mu$ are obtained from the diagonalization of the matrix:
\begin{eqnarray}
    H_{(n,\kappa),(n',\kappa')}^\mu &=& \mem{\phi_{n,\kappa,\mu}}{\hat{H}}{\phi_{n',\kappa',\mu}}
    \nonumber \\[0.2cm]
    && \hspace*{-2cm} = \epsilon_{n,\kappa} \, \delta_{n,n'} \delta_{\kappa,\kappa'} \nonumber \\
    && \hspace*{-2cm} + \mem{\phi_{n,\kappa,\mu}}{\sum\limits_{l=1}^{l_{max}} V_l(r) P_l(\cos \theta)}{\phi_{n',\kappa',\mu}} \, .
\end{eqnarray}
Here we expanded the two--center potential (\ref{eq:potential_expansion}) into monopole and higher--multipole terms, and used the fact that ${\hat H}_0 \, \phi_{n,\kappa,\mu}(\bfr) = \epsilon_{n,\kappa} \, \phi_{n,\kappa,\mu}(\bfr)$ with ${\hat H}_0$ being the ``monopole'' Hamiltonian.

For the further theoretical analysis it will be convenient to represent the eigensolutions (\ref{eq:expansion_PSI}) of the full two--center Hamiltonian as the sum of their \textit{partial--wave} contributions:
\begin{equation}
    \label{eq:wave_function_decomposition}
    \Phi_{N,\mu}(\bfr) = \sum\limits_{\kappa} \Phi_{N,\mu}^\kappa(\bfr) \, ,
\end{equation}
where:
\begin{equation}
    \Phi_{N,\mu}^\kappa(\bfr)=\sum\limits_n A_{n,\kappa}^{N,\mu} \; \phi_{n,\kappa,\mu}(\bfr)\,.
\end{equation}
These partial--wave contributions posses a well--defined symmetry and, hence, can be written in the standard form:
\begin{equation}
    \label{eq:partial_wave_function}
    \Phi_{N,\mu}^\kappa(\bfr)=\left(\begin{array}{r} g_{N,\mu}^\kappa(r)
    \Omega_{\kappa,\mu}(\theta,\phi)\\
    i\,f_{N,\mu}^\kappa(r)
    \Omega_{-\kappa,\mu}(\theta,\phi)\end{array}
    \right) \, ,
\end{equation}
where $g(r)$ and $f(r)$ are the large and small radial components and $\Omega_{\pm\kappa, \mu}(\theta,\phi)$ is the Dirac spinor.

Having derived the eigenfunctions and eigenenergies of the two--center Dirac Hamiltonian (\ref{eq:Dirac}), we are ready to generate the corresponding Green's function. This function is obtained by the summation over all quasimolecular states $\ketm{N, \mu}$, that are characterized by the number $N$ and by the projection $\mu$ of the total angular momentum:
\beqn
    \label{spectral}
    G(\bfx,\bfy;\omega)=\sum\limits_{N,\mu}
    \frac{\Phi_{N,\mu}(\bfx)\Phi_{N,\mu}^+(\bfy)}{\omega-{\mathcal E}_{N,\mu} +
    i0({\mathcal E}_{N,\mu}+1)}\,.
\eeqn
One can note, that the positions of the poles of the Green's function are changed in comparison with the conventional definition $\omega - {\mathcal E}_{N, \mu}(1-i0)$. This is done to account for the fact that one--electron states with the energy in the region $-1 < {\cal E}_{N, \mu} < 0$ are electronic (and not positronic) ones and should be treated as the positive--energy states \cite{gyulassy:75}. Since in the present study we will not discuss the over--critical regime in which bound molecular states can lie in the negative continuum, this definition of the poles is justified.

\subsection{QED corrections to energy levels}
\label{subsec:QED_corrections}

\begin{figure}[t]
\centering
\includegraphics[clip=true,width=0.15\columnwidth]{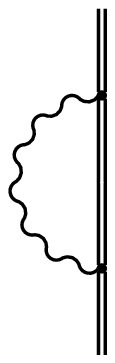}
\includegraphics[clip=true,width=0.28\columnwidth]{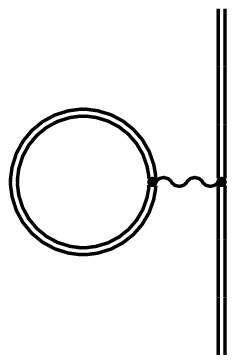}
\caption{The Feynman diagrams representing self--energy (left) and vacuum--polarization (right) corrections to the energy levels.}
\label{Fig2}
\end{figure}

In the previous section we have briefly discussed how to generate the wave and Green's functions for an electron, moving in the field of two nuclei. Now we are ready to employ these functions for the evaluation of the QED corrections to the quasimolecular energy levels. To the first order in $\alpha$ these corrections arise due to the self--energy (SE) and vacuum polarization (VP) effects, with the corresponding Feynman diagrams displayed in Fig.~\ref{Fig2}. In what follows, we will derive the SE and VP corrections to the energy ${\mathcal E}_{a} \equiv {\mathcal E}_{N_a, \mu_a}$ of a particular molecular state state $\ketm{a} = \ketm{N_a, \mu_a}$, characterized by the angular momentum projection $\mu_a$. For the sake of brevity, we will use below the short--hand notations for the wave function of this state, $\Phi_{a}(\bfr) \equiv \Phi_{N_a,\mu_a}(\bfr)$, and for its partial--wave contributions $\Phi_{a}^\kappa(\bfr) \equiv \Phi_{N_a,\mu_a}^\kappa(\bfr)$.

\subsubsection{Self energy}

\begin{figure}[t]
\centering
\includegraphics[clip=true,width=0.9\columnwidth]{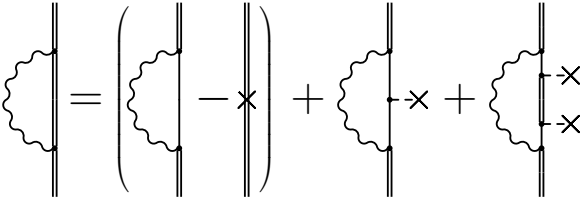}
\caption{The decomposition of self--energy diagram into zero--, first--, and many--potential terms.}
\label{seeq}
\end{figure}

We start the evaluation of the first--order QED corrections from the self--energy term. Here we will follow the standard approach, discussed in detail in Refs.~\cite{blundell:91:b,yerokhin:99:pra}.
In this approach, the internal electron propagator, displayed in the left--hand side of Fig.~\ref{seeq} by the double line, is expanded in powers of the interaction with external (nuclear) field. The first term of this expansion is known as the zero--potential one and contains free--electron propagator. In order to obtain the finite result, this zero--order term has to be covariantly regularized and evaluated together with the counter--term of the mass renormalization. The second term in the external--field expansion is known as the vertex or one--potential term. It contains the free--electron vertex operator and must be regularized before the evaluation, as discussed in \cite{yerokhin:99:pra}. Finally, the last term in Fig.~\ref{seeq} is known as many--potential term and does not require any regularization. In what follows, we briefly discuss the evaluation of the energy corrections $\Delta E_{\rm SE}^0$, $\Delta E_{\rm SE}^1$ and $\Delta E_{\rm SE}^{\rm many}$ that arise from these three terms.

The regularized zero-- and one--potential terms are conveniently calculated in
the momentum space. Using notations from Ref. \cite{yerokhin:99:pra}, we can write the corresponding energy shifts as:
\beqn \label{se_0}
\Delta E_{\rm SE}^0&=&\sum\limits_{\kappa_1,\kappa_2=-K}^K \int \frac{d^3\bfp}{(2\pi)^3}
\bar{\Phi}^{\kappa_1}_a(\bfp)
\Sigma_{\rm R}^{(0)}(\bfp) {\Phi}_a^{\kappa_2}(\bfp)\,,\\\label{se_1}
\Delta E_{\rm SE}^1&=&\sum\limits_{\kappa_1,\kappa_2=-K}^K\int
\frac{d^3\bfp_1}{(2\pi)^3}\int \frac{d^3\bfp_2}{(2\pi)^3}
\bar{\Phi}_a^{\kappa_1}(\bfp_1) \Gamma_{\rm R}^0(\bfp_1,\bfp_2)\nonumber
\\ &\times& V(|\bfp_1-\bfp_2|){\Phi}_a^{\kappa_2}(\bfp_2) \, ,
\eeqn
where $\Sigma_{\rm R}^{(0)}$ and $\Gamma_{\rm  R}^0$ are the operators arising from the perturbation expansion of the bound--electron Green function in powers of external potential. The explicit form of these operators and further details can be found in Ref.~\cite{yerokhin:99:pra}. The evaluation of the energy corrections $\Delta E_{\rm SE}^0$ and $\Delta E_{\rm SE}^1$ in momentum space requires, moreover, the knowledge of the Fourier transforms of the two--center potential:
\begin{equation}
    \label{Vmom}
    V(\bfp) = \exp(-i \bfp \cdot \bfR)V_1(p) + \exp(i \bfp \cdot \bfR)V_2(p) \, ,
\end{equation}
with $V_i(p) = -\alpha Z_i \int_{-\infty}^{+\infty} {\rm d}r \, {\rm e}^{-i\bfp \cdot \bfr} \, (1 + y_i(r))/r$, and of the quasimolecular wave function:
\beqn
    \label{Psimom}
    \Phi^\kappa_a(\bfp)&=&i^{-l}\left(\begin{array}{r}
    \tilde{g}_a^\kappa(p) \, \Omega_{\kappa,\mu_a}(\theta_p,\phi_p)\\
    \tilde{f}_a^\kappa(p) \, \Omega_{-\kappa,\mu_a}(\theta_p,\phi_p)\end{array}\right)
    \,.
\eeqn
In the latter expression, $\tilde{g}^\kappa_a(p) \equiv \tilde{g}^\kappa_{N_a, \mu_a}(p)$ and $\tilde{f}^\kappa_a(p) \equiv \tilde{f}^\kappa_{N_a, \mu_a}(p)$ are the Fourier transforms of large and small components of the partial wave function (\ref{eq:partial_wave_function}).

By inserting wave function (\ref{Psimom}) into Eq.~(\ref{se_0}) and performing the angular integration, we obtain the zero--potential correction:
\beqn
\Delta E_{\rm SE}^0&=&\frac{\alpha}{4\pi}\frac{1}{(2\pi)^3}
\sum\limits_{\kappa}
\int\limits_0^\infty p^2 dp
\left[a(\rho)(\tilde{g}_a^\kappa\tilde{g}_a^\kappa-\tilde{f}_a^\kappa\tilde{f}_a^\kappa)\right.
\nonumber \\ &&
+\left. b(\rho)({\mathcal E}_a(\tilde{g}_a^\kappa\tilde{g}_a^\kappa+\tilde{f}_a^\kappa\tilde{f}_a^\kappa)
+2p\tilde{g}_a^\kappa\tilde{f}_a^\kappa\right] \, ,
\eeqn
which contains only diagonal terms $\kappa_1 = \kappa_2$ in the partial--wave summation. Here, moreover, $a(\rho)$ and $b(\rho)$ are the components of the free self--energy function whose explicit form is given in Ref.~\cite{yerokhin:99:pra}.

The evaluation of the one--potential energy correction $\Delta E_{\rm SE}^1$ is a bit more complicated and requires a multipole expansion of the two--center potential $V(\bfp_1-\bfp_2)$. By using in Eq.~(\ref{Vmom}) the well--known decomposition of the exponential function
\beq
\exp(i \bfp \cdot \bfr)=4 \pi \sum\limits_{lm} i^l j_l(pr) Y^*_{lm}(\theta_p,\phi_p)
Y_{lm}(\theta_r,\phi_r)
\eeq
and performing the angular integration in (\ref{se_1}) one can obtain, after some algebra
\begin{widetext}
\beqn
\Delta E_{\rm
  SE}^1&=&\frac{\alpha}{(2\pi)^5}\sum\limits_{\kappa,\kappa',l,l',L}\int\limits_0^\infty
dp'\int\limits_0^\infty dp\,
p^2p'^2
\left[U_1^{l,l',L,\kappa',\kappa}(p,p')D_{l',l,L,\kappa',\kappa}
  +U_2^{l,l',L,\kappa',\kappa}(p,p')D_{l',l,L,-\kappa',-\kappa}\right]\,,
\eeqn
where angular and radial functions read as:
\beqn
D_{l',l,L,\kappa',\kappa}&=&i^{l_{\kappa'}-l_\kappa+l'-l}C^{l_{\kappa'}0}_{l'0;L0}C^{l_{\kappa}0}_{l0;L0}
\frac{(2L+1)(2l'+1)(2l+1)}
{4\pi\sqrt{(2l_{\kappa'}+1)(2l_\kappa+1)}}
\nonumber\\ &&\times
\sum\limits_M\left[
C^{j_{\kappa'}1/2}_{l_{\kappa'}M;1/2\mu_a-M}C^{j_{\kappa}1/2}_{l_{\kappa}M;1/2\mu_a-M}
C^{l_{\kappa'}M}_{l'0;LM}C^{l_{\kappa}M}_{l0;LM}\right]\,,\\
U_i^{l,l',L,\kappa',\kappa}(p,p')&=&\int\limits_{-1}^1 d\xi {\cal
  F}_i^{\kappa'\kappa}(p,p',\xi)f_{ll'}(p,p',\xi)P_L(\xi)\,,\\
f_{ll'}(p,p',\xi)&=&j_{l'}(p'R)j_l(pR)V_1(p_{12})
+(-1)^{(l+l')}j_{l'}(p'R)j_l(pR)V_2(p_{12})\,.
\eeqn
\end{widetext}
In the above expressions, $p_{12}=\sqrt{p^2+p'^2-2pp'\xi}$ is the absolute value of the momenta difference, $\xi = \bfp \cdot \bfp'/ pp'$, $j_l$ is the spherical Bessel function, and $C^{JM}_{j_1m_1;j_2m_2}$ is Clebsch--Gordan coefficient. Finally, the functions ${\cal F}_i^{\kappa'\kappa}$ represent the components of the free vertex operator sandwiched between the radial wave function components $\tilde{g}$ and $\tilde{f}$, with the explicit formulas given in Appendix A of Ref.~\cite{yerokhin:99b}.


While the diverging zero-- and one--potential self--energy corrections were evaluated in the momentum space, the remaining many--potential term does not require renormalization and can be calculated in coordinate representation. Its formal expression is given by:
\begin{widetext}
\beqn\label{Emany}
\Delta E^{\rm many}_{\rm SE}&=&2i\alpha\int\limits_{-\infty}^\infty d
\omega \int d^3\bfr_1 \int d^3 \bfr_2 \int d^3 \bfr_3 \int d^3 \bfr_4\nonumber \\ &&\times
\Phi_a(\bfr_1)^+F({\mathcal E}_a-\omega;\bfr_1,\bfr_2)V(\bfr_2)G({\mathcal E}_a-\omega;\bfr_2,\bfr_3)
V(\bfr_3)F({\mathcal E}_a-\omega;\bfr_3,\bfr_4)I(\omega,|\bfr_1-\bfr_4|)\Phi_a(\bfr_4)\,,
\eeqn
\end{widetext}
where $F$ is the one--electron Green function in the absence of external field, and the operator $I(\omega)$ is defined as
\beq\label{eq:propagator}
I(\omega,|\bfr_1-\bfr_2|)=(1-\balpha_1 \cdot \balpha_2)\frac{\exp (i|\omega| |\bfr_1-\bfr_2|)}{ |\bfr_1-\bfr_2|}\,,
\eeq
with $\balpha$ being the vector of Dirac matrices.

By making use of the spectral representation for the free--electron-- $F$ and full--potential-- $G$ Green functions, and by introducing the auxiliary function
\beq
\tilde{\psi}_{n,\mu}(\omega)=\sum\limits_k \frac{\langle u_k|V|\psi_{n,\mu}\rangle u_k}
{{\mathcal E}_a-\omega-\epsilon_k(1-i0)}
\eeq
one can obtain the many--potential correction (\ref{Emany}) in the form:
\beqn \label{dEmany}
\Delta E^{\rm many}_{\rm SE} = 2i\alpha\int\limits_{-\infty}^\infty {\rm d}
\omega \sum\limits_{n,\mu} \frac{\langle \Phi_a \tilde{\psi}_{n,\mu}(\omega)|I(\omega)|
\tilde{\psi}_{n,\mu}(\omega)
\Phi_a
\rangle}{{\mathcal E}_a-\omega-{\mathcal E}_{n,\mu}+i0({\mathcal E}_{n,\mu}+1)}\,.\nonumber \\
\eeqn
Here, $u_k$ and $\epsilon_k$ are the eigenvectors and eigennumbers of free (without external potential) Dirac equation.

In order to evaluate $\Delta E^{\rm many}_{\rm SE}$ it is convenient to rotate the integration contour over $\omega$ to the imaginary axis. In this case, the oscillatory behavior of $I(\omega)$ changes to the exponential damping, which results in the fast convergence of the integral over $\omega$. Moreover, this rotation leads to the appearance of the pole contributions arising when the
integration contour crosses the pole $\omega={\mathcal E}_a-{\mathcal E}_{n,\mu}+i0({\mathcal E}_{n,\mu}+1)$. The many--potential correction can be written, therefore, as a sum:
\begin{equation}
    \label{eq:many-potential-final}
    \Delta E^{\rm many}_{\rm SE} = \Delta E_{\rm pole} + \Delta E_{\rm int} \,
\end{equation}
of the pole term:
\begin{widetext}
\begin{equation}
    \Delta E_{\rm  pole} = 4\pi \alpha\sum\limits_{{\mathcal E}_a\ge{\mathcal E}_{n,\mu}>-1}
    \left( 1-\frac{\delta_{{\mathcal E}_a{\mathcal E}_{n,\mu}}}{2} \right) 
    \mem{\Phi_a \tilde{\psi}_{n,\mu}({\mathcal E}_a-{\mathcal E}_{n,\mu})}
    {I({\mathcal E}_a-{\mathcal E}_{n,\mu})}
    {\tilde{\psi}_{n,\mu}({\mathcal E}_a-{\mathcal E}_{n,\mu})\Phi_a} \, ,
\end{equation}
and of the integral term:
\begin{equation}
    \Delta E_{\rm int} = -4\alpha \, {\rm Re}
    \int\limits_0^\infty {\rm d}w \sum\limits_{n,\mu}
    \frac{\langle \Phi_a \tilde{\psi}_{n,\mu}(iw)|\tilde{I}(w)|
    \tilde{\psi}_{n,\mu}(iw) \Phi_a \rangle}{{\mathcal E}_a-iw-{\mathcal E}_{n,\mu}}\,
\end{equation}
\end{widetext}
and where the operator (\ref{eq:propagator}) after the contour rotation is given by:
\begin{equation}
    \tilde{I}(w) = (1-\balpha_1 \cdot \balpha_2) \, \frac{\exp(-w \left| \bfr_1-\bfr_2 \right|)}{\left| \bfr_1-\bfr_2 \right|}\,.
\end{equation}
%
%

%
%
%
%
\subsection{Vacuum polarization}

\begin{figure}[b]
\centering
\includegraphics[clip=true,width=0.9\columnwidth]{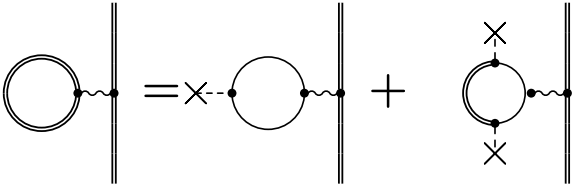}
\caption{The diagram equation for the calculation of vacuum polarization correction to quasimolecular energy levels. Two diagrams on the right--hand side represent Uehling and Wichmann-Kroll contributions.}
\label{vpeq}
\end{figure}

Unlike the self--energy effect, the vacuum polarization contribution to a quasimolecular energy level
\beq\label{UVP}
\Delta E_{\rm VP}=\langle \Phi_a|U_{\rm VP}|\Phi_a \rangle\,,
\eeq
can be obtained as an expectation value of the \textit{local} VP potential:
\beq
U_{\rm VP}(\bfr_1)=\frac{\alpha}{2\pi i}\int\,{\rm d}^3
\bfr_2\int\limits_{-\infty}^\infty {\rm d}\omega\frac{{\rm Tr} [G(\omega;\bfr_2,\bfr_2)]}{\left| \bfr_1-\bfr_2 \right|}\,,
\eeq
where $G(\omega;\bfr_2,\bfr_2)$ denotes the one--electron Green function (\ref{spectral}). For the evaluation of the VP correction $\Delta E_{\rm VP}$ is usually separated into two parts, known as the Uehling and Wichmann--Kroll contributions. The Feynman diagrams for these two contributions are presented in Fig.~\ref{vpeq}. The leading, Uehling, contribution \cite{uehling:35} is divergent and requires charge renormalization. The renormalized expression for the Uehling potential is well known for a single nucleus:
\beqn\label{UUe}
U_{\rm Ue}(Z, r) &=& -\frac 83 \alpha^2 Z \int_0^\infty dr' r'  \rho_n(r')
\nonumber \\[0.2cm]
              & & \hspace*{-0.8cm} \times \:
                  \int_1^\infty dt \left(1+\frac{1}{2t^2}\right) \:
                  \frac{\sqrt{t^2-1}}{t^2}  \nonumber \\[0.2cm]
              & & \hspace*{-0.8cm} \times \:
                  \frac{\exp{(-2|r-r'|t)}-\exp{(-2(r+r')t)}}{4rt}\, ,
\eeqn
with $\rho_n(r')$ being the nuclear charge distribution. By making use of this expression, one can calculate the Uehling correction to the energy level $\ketm{a}$ for the two--center potential as:
\beqn
    \Delta E_{\rm Ue}=\mem{\Phi_a}{U_{\rm Ue}(Z_1, r_1) + U_{\rm Ue}(Z_2, r_2)}{\Phi_a} \, ,
\eeqn
where $r_i = \left|\bfr - \bfR_i \right|$. The numerical evaluation of $\Delta E_{\rm Ue}$ is performed most conveniently if the potentials $U_{\rm Ue}(Z_1, r_1)$ are expanded in terms of Legendre polynomials, similar to Eq.~(\ref{eq:potential_expansion}).

The remaining part of the VP potential, known as the Wichmann--Kroll term \cite{wichmann:56}, can be  written as:
\beqn\label{eq:Uwk}
U_{\rm WK}(\bfr_1)&=&\frac{\alpha}{2 \pi i}\int \frac{{\rm d}^3 \bfr_2}{|\bfr_1-\bfr_2|}
\nonumber \\
&&\times\int\limits_{-\infty}^{\infty}{\rm d}\omega {\rm
  Tr}\left[G^{(2+)}(\omega;\bfr_1,\bfr_2)\right] \, , \\
G^{(2+)}(\omega;\bfr_1,\bfr_2)&=&\int {\rm d}^3\bfx \int {\rm d}^3\bfy
F(\omega;\bfr_1,\bfx) V_{2c}(\bfx)\nonumber \\
&&\times G(\omega;\bfx,\bfy)V_{2c}(\bfy)F(\omega;\bfy,\bfr_2)\, ,
\eeqn
where $V_{2c}(\bfx)$ is two--center potential (\ref{eq:potential_expansion}). As seen from these expressions, the evaluation of the Wichmann--Kroll correction can can be traced back to the Green functions of a free electron, $F$, and of an electron, moving in a two--center potential, $G$. For the latter, one can use again the  spectral representation, see Eq.~(\ref{spectral}). However, due to the strong cancellation between the contributions from positive-- and negative--energy states, this approach requires enormously large number of basis functions $\Phi_{n,\mu}$. In order to accelerate the numerical procedure, we employ algorithm, proposed in Ref.~\cite{mohr:98:rev}, and compute $G$ within the monopole approximation. Together with the analytic representation of the free--electron function $F$ in terms of spherical Bessel and Hankel functions \cite{mohr:74:a,mohr:74:b}, this (monopole) approximation allows fast and accurate computation
of the charge density corresponding to the Wichmann--Kroll potential:
\beqn\label{rhowk}
    \rho_{wk}(\bfr_1) = \frac{\alpha}{2 \pi i}\int
    {\rm d}^3\bfr_2\int\limits_{-\infty}^{\infty} {\rm d}\omega \;
    {\rm Tr}\left[G^{(2+)}(\omega;\bfr_1,\bfr_2)\right]\,.
\eeqn
The integral over the energy $\omega$ circulating in the electron loop can be accurately evaluated upon the rotation of the integration contour to the imaginary axis:
\beqn
    \label{rhowk_rot}
    \rho_{wk}(\bfr_1)&=&\frac{\alpha}{\pi}\int
    {\rm d}^3\bfr_2\int\limits_{0}^{\infty} {\rm d}w \;
    {\rm Tr}\left[G^{(2+)}(iw;\bfr_1,\bfr_2)\right]
    \nonumber\\ &-& \alpha\sum\limits_{-1<{\mathcal E}_{N,\mu} \le 0}
    \left(1-\frac{\delta_{{\mathcal E}_{N,\mu},0}}{2}\right)|\Phi_{N, \mu}(\bfr_1)|^2\,.
\eeqn
Here, the last term appears only in strong fields, where one or more bound molecular states have negative energies, ${\mathcal E}_{N,\mu} < 0$. Such states provide the poles of the Green function (\ref{spectral}) with negative real and imaginary parts. These poles will be crossed during the rotation of the contour in the complex plane and their contribution must be compensated by the corresponding pole term. The evaluation of the residual in the point $\omega = {\mathcal E}_a$ leads to the simple formula for the pole contribution with the charge density of bound state.

%
%
\begin{table*}
\caption{The zero--order energy ${\mathcal E}_a = {\mathcal E}_{1\sigma_g}$ and the first--order QED corrections for the $1\sigma_g$ ground state of U$_2^{183+}$ quasimolecule. The units are eV and the normalized relativistic units, defined by Eq.~(\ref{eq:normalized_energy}), in the upper and lower part of the table, respectively. \label{tabU2}}
\begin{ruledtabular}
\begin{tabular}{lllll}
Distance [fm]&
\multicolumn{1}{c}{${\mathcal E}_a$}&
\multicolumn{1}{c}{$\Delta E_{\rm SE}$ } &
\multicolumn{1}{c}{$\Delta E_{\rm VP}$ } &
\multicolumn{1}{c}{$\Delta E_{\rm QED}$ }
\\[0.1cm] \colrule
\\[-0.15cm]
  40 &$-4.62554(96)\times 10^{   5}$&$6.755(98)\times 10^{   3}$&$-4.720(69)\times 10^{   3}$&$2.03(12)\times 10^{   3}$\\
  50 &$-3.8862(43)\times 10^{   5}$&$5.7285(33)\times 10^{   3}$&$-3.6830(21)\times 10^{   3}$&$2.0454(39)\times 10^{   3}$\\
  80 &$-2.5398(42)\times 10^{   5}$&$3.974(13)\times 10^{   3}$&$-2.0680(72)\times 10^{   3}$&$1.906(15)\times 10^{   3}$\\
 100 &$-1.985(14)\times 10^{   5}$&$3.317(36)\times 10^{   3}$&$-1.571(17)\times 10^{   3}$&$1.745(40)\times 10^{   3}$\\
 200 &$-4.90(17)\times 10^{   4}$&$1.821(36)\times 10^{   3}$&$-6.63(13)\times 10^{   2}$&$1.158(38)\times 10^{   3}$\\
 250 &$-6.11(18)\times 10^{   3}$&$1.480(18)\times 10^{   3}$&$-5.011(61)\times 10^{   2}$&$9.78(19)\times 10^{   2}$\\
 500 &$1.1620(80)\times 10^{   5}$&$7.36(31)\times 10^{   2}$&$-2.120(90)\times 10^{   2}$&$5.24(32)\times 10^{   2}$\\
 700 &$1.698(12)\times 10^{   5}$&$5.17(35)\times 10^{   2}$&$-1.434(99)\times 10^{   2}$&$3.73(37)\times 10^{   2}$\\
1000 &$2.219(20)\times 10^{   5}$&$3.58(39)\times 10^{   2}$&$-9.9(10)\times 10^{   1}$&$2.59(40)\times 10^{   2}$\\ \\
\hline \\
  40 &$-36.2078(75)$&$0.5297(77)$&$-0.3694(54)$&$0.1592(94)$\\
  50 &$-38.026(42)$&$0.56052(31)$&$-0.36048(20)$&$0.20014(37)$\\
  80 &$-39.762(66)$&$0.6221(21)$&$-0.3247(11)$&$0.2984(24)$\\
 100 &$-38.84(38)$&$0.6491(71)$&$-0.3086(33)$&$0.3425(79)$\\
 200 &$-19.20(79)$&$0.713(14)$&$-0.2606(51)$&$0.453(14)$\\
 250 &$-2.990(89)$&$0.7240(88)$&$-0.2451(29)$&$0.4798(93)$\\
 500 &$113.70(89)$&$0.720(30)$&$-0.2074(87)$&$0.513(31)$\\
 700 &$232.7(17)$&$0.718(48)$&$-0.206(13)$&$0.511(50)$\\
1000 &$434.3(40)$&$0.702(76)$&$-0.194(21)$&$0.517(79)$\\
\end{tabular}
\end{ruledtabular}
\end{table*}
%
%

The monopole approximation for the evaluation of the Wichmann--Kroll correction to quasimolecular levels is well justified for small internuclear distances $R$, where $V_{2c}(\bfr) \approx V_0(r)$. This is not the case for large $R$'s, for which higher--multipole terms in the two--center potential play an essential role. For large distances $R$, however, the QED correction $\Delta E_{\rm QED} = \Delta E_{\rm SE} + \Delta E_{\rm VP}$ approaches its value for a single atom, for which the Wichmann--Kroll contribution to $\Delta E_{\rm QED}$ does not exceed 1--2 \% \cite{soff:88:vp,manakov:89:zhetp}. This is much smaller than the numerical uncertainty of our results for large $R$, that is estimated to be 10-15\%, as will be shown in the next Section. We can conclude, therefore, that the truncation of a full multipole expansion for the Wichmann--Kroll contribution to the single monopole term is justified for all internuclear distances.

%
%
\section{Results and discussion}
\label{result}

The theoretical approach, outlined above, can be used to calculate the QED corrections to energy levels for an arbitrary diatomic quasimolecule. In the present study, we discuss calculations for the ground state $\ketm{a} = \ketm{1\sigma_g}$ of the ${\rm U}_2^{183+}$, ${\rm U}\,{\rm Pb}^{173+}$ and ${\rm Bi}\,{\rm Au}^{161+}$ dimers. This choice of quasimolecules allows us to investigate the QED corrections both for homo-- and heteronuclear cases. The latter case attracts also a particular interest because of experiments in which the formation of heteronuclear quasimolecules was observed in ion--atom collisions \cite{verma:05,verma:06:rphys,verma:06:nimb}. Even though thus produced dimers contain many electrons, our calculations may be relevant for these and similar experiments, because the many--electron effects are not so important for the low--lying molecular levels.

Before we present the results of our calculations, let us briefly recall the details of the
numerical approach. In the first step, we employed the basis of 70 radial $B$--splines in order to generate the eigensolutions $\phi_{n, \kappa, \mu}$ of the \textit{monopole} Dirac Hamiltonian with the angular quantum number in the range $-25 \leq \kappa \leq 25$. These monopole solutions form $50 - 2|\mu| + 1$  partial contributions to represent the wave functions $\Phi_{N, \mu}(\bfr)$ of the full (two--center) Hamiltonian, see Eq.~(\ref{eq:expansion_PSI}). The investigation of the convergence of the many--potential term for the self--energy correction (\ref{dEmany}) demonstrated, that inclusion of all states with $|\mu|\leq 5$ into the spectral representation (\ref{spectral}) leads to the relative accuracy better than $10^{-3}$. This is definitely less than the uncertainty, introduced by the truncation of the expansions (\ref{eq:potential_expansion}) and (\ref{eq:expansion_PSI}) at large value $l_{max}$ or $K$.

In comparison to the method described in our previous paper \cite{artemyev:15} we have made a minor change in the present algorithm, distributing the knots on the radial grid of B splines. Namely, in Ref.~\cite{artemyev:15} we used the predefined number of radial knots in the inner (between two nuclei) and outer regions. In the present approach we first find the radial knot distribution which minimizes the ground state energy and only then perform the QED calculations. This explains slight difference between the present and previous predictions for the homonuclear case U$_2^{183+}$. This difference, however, does not exceed 3\% for all internuclear distances, thus confirming the robustness of our calculations.

Another (technical) difference from the previous study \cite{artemyev:15}, is a novel approach to estimate the uncertainty of our predictions. The main source of this uncertainty is the \textit{truncation} of the multipole expansions of molecular wave functions (\ref{eq:expansion_PSI}) and potential (\ref{eq:potential_expansion}), which is generic problem of two--center calculations, performed in spherical coordinates. In order to assess errors introduced by this truncation, we compared the (zero--order) ground--state energy ${\mathcal E}_a = {\mathcal E}_{1\sigma_g}$ and the VP Uehling correction $\Delta E_{\rm Ue}$, obtained in the present work, with the predictions, based on the solution of the two--center Dirac equation in Cassini coordinates. The Cassini coordinate approach to the description of quasimolecular structure has been discussed by us in Ref.~\cite{artemyev:10} and has been shown to be free of ``truncation problem'', providing closed expressions for the potential and wave functions. In the past, we successfully employed this approach for the computation of the ${\mathcal E}_a$ and $\Delta E_{\rm Ue}$ for any internuclear distance and with the relative accuracy below $10^{-6}$. However, due to absence of efficient algorithms for the evaluation of the many--dimensional integrals and Fourier transforms, the method of Cassini coordinates has not been implemented so far for the evaluation of SE and Wichmann-–Kroll VP corrections. In our study, therefore, we use the known high-precision results for the ${\mathcal E}_a$ and  $\Delta E_{\rm Ue}$ in Cassini coordinates to estimate their uncertainties in the spherical basis.

Having briefly discussed the numerical details and uncertainty analysis, we are ready to present results of our calculations. First, we revisit the homonuclear case of the U$_2^{183+}$ dimer, which has been studied already in our previous work \cite{artemyev:15}. For the ground 1$\sigma_g$ state of U$_2^{183+}$, we present in the upper part of Table~\ref{tabU2} the zero--order energy $ {\cal E}_a$, the self--energy $\Delta E_{\rm SE}$ and vacuum polarization $\Delta E_{\rm VP}$ corrections, as well as their sum $\Delta E_{\rm QED} = \Delta E_{\rm SE} + \Delta E_{\rm VP}$. The energies are given here as a function of the inter--nuclear distance $R$. We performed calculations for distances, ranging from $R$~=~40~fm, for which nuclei come very close to each other and molecular effects become of paramount importance, up to $R$~=~1000~fm, where the energy spectrum starts to resemble that of a single U$^{91+}$ ion. As seen from the table, both the zero--order energy ${\mathcal E}_a$ and QED corrections vary significantly with $R$. At small distances, for example, the energy ${\mathcal E}_a$ is nearing the value of $-mc^2 \approx - $~511~keV, thus indicating that the ground 1$\sigma_g$ state almost reaches the negative continuum threshold. The sum of SE and VP contributions for this sub--critical regime is about 2~keV, which implies 0.4~\% QED correction to ${\mathcal E}_a$. The ground--state energy ${\mathcal E}_a$ increases with the inter--nuclear distance and is about $E_0 \approx 2 \times 10^5$~eV for $R = 1000$~fm. For this---rather large---$R$, the sum of QED corrections reaches the value of $\Delta E_{\rm QED} = 259 \pm 40$~eV, and becomes comparable to the atomic (single--center) result $\Delta E_{\rm QED}({\rm U}^{91+}) = 266.45$~eV \cite{yerokhin:15}.

When comparing predictions from Table~\ref{tabU2} with those from our previous study \cite{artemyev:15}, one can note $\approx 1$\% difference in results for the correction $\Delta E_{\rm QED}$. As mentioned already above, this is due to the modified radial basis, used in the present work. This few--percent difference is much below the accuracy of our calculations, which allows us to state a good agreement with the previous results.

To better understand the behaviour of the QED corrections as the inter--nuclear distance $R$ changes, one can evaluate the \textit{normalized} energy quantities:
\begin{equation}
    \label{eq:normalized_energy}
    \tilde{E} = E [{\rm eV}] \times R [{\rm fm}] \times \frac{\alpha^2}{2{\rm Ry}} \, ,
\end{equation}
where {\rm Ry} is the Rydberg energy. In the lower part of Table~\ref{tabU2}, we present the normalized zero--order energy as well as the QED corrections. As one can see, both $\widetilde{\Delta E}_{\rm SE}$ and $\widetilde{\Delta E}_{\rm VP}$ are of the order of unity and weakly depend on $R$. It implies that for the inter--nuclear distances, considered in our study, the QED corrections scale approximately as $1/R$. This scaling, however, does not hold for larger distances, $R \gg 1000$~fm, where both $\Delta E_{\rm SE}$ and $\Delta E_{\rm VP}$ reach their ``atomic values'' that are independent on $R$.

In order to investigate the application of our approach to a heteronuclear case, we have performed calculations for ${\rm U} \, {\rm Pb}^{173+}$ and ${\rm Bi} \, {\rm Au}^{161+}$ quasimolecules. In Tables \ref{tabUPb} and \ref{tabBiAu} the energy ${\mathcal E}_a$ of the ground 1$\sigma_g$ state of these dimers as well as the QED corrections are presented again as functions of inter--nuclear distance $R$. One can see that both ${\mathcal E}_a(R)$ and $\Delta E_{\rm QED}(R)$ behave \textit{qualitatively} similar to that was predicted for the U$_2^{183+}$ case. That is, the electron is most strongly bound for small inter--nuclear distances, for which, however, ${\mathcal E}_a$ is still rather far from the negative continuum threshold, as can be expected for ``lighter'' dimers ${\rm U} \, {\rm Pb}^{173+}$ and ${\rm Bi} \, {\rm Au}^{161+}$. The predictions for large $R$ again resemble atomic calculations: in the heteronuclear case both ${\mathcal E}_a$ and $\Delta E_{\rm QED}$ approach results obtained for an isolated heaviest nucleus.

Tables~\ref{tabU2}--\ref{tabBiAu} indicate that the relative accuracy of the QED calculations both for hetero-- and homonuclear dimers varies from $\approx 0.1$\% for the small distances to $\approx 10$\% for the large ones. This behaviour can be well understood from the fact that the convergence of multipole expansion of molecular wave functions (\ref{eq:expansion_PSI}) deteriorates as the distance between nuclei increases. Nevertheless, we argue that even a rather moderate basis of 70 B-splines and 50 different $\kappa$'s, employed in this paper, provides a good accuracy of the QED predictions at rather large distances up to $R \approx$~1000~fm. This proves the applicability of the developed approach for the quantum chemistry purposes.

\begin{widetext}

\begin{table*}
\caption{The same as Table~\ref{tabU2} but for the ${\rm U} \, {\rm Pb}^{173+}$ quasimolecule.\label{tabUPb}}
\begin{ruledtabular}
\begin{tabular}{lllll}
Distance [fm]&
\multicolumn{1}{c}{${\mathcal E}_a$}&
\multicolumn{1}{c}{$\Delta E_{\rm SE}$}&
 \multicolumn{1}{c}{$\Delta E_{\rm VP}$} &
 \multicolumn{1}{c}{$\Delta E_{\rm QED}$} \\[0.1cm] \colrule
 \\[-0.15cm]
  25 &$-4.006(12)\times 10^{   5}$&$7.41(21)\times 10^{   3}$&$-6.00(17)\times 10^{   3}$&$1.40(27)\times 10^{   3}$\\
  50 &$-2.35761(91)\times 10^{   5}$&$4.724(71)\times 10^{   3}$&$-2.501(37)\times 10^{   3}$&$2.223(80)\times 10^{   3}$\\
 100 &$-1.000(10)\times 10^{   5}$&$2.819(60)\times 10^{   3}$&$-9.97(21)\times 10^{   2}$&$1.821(64)\times 10^{   3}$\\
 300 &$7.856(76)\times 10^{   4}$&$1.102(34)\times 10^{   3}$&$-3.28(10)\times 10^{   2}$&$7.73(36)\times 10^{   2}$\\
 500 &$1.5361(64)\times 10^{   5}$&$6.66(40)\times 10^{   2}$&$-1.76(10)\times 10^{   2}$&$4.89(42)\times 10^{   2}$\\
 700 &$2.001(10)\times 10^{   5}$&$4.72(40)\times 10^{   2}$&$-1.20(10)\times 10^{   2}$&$3.52(42)\times 10^{   2}$\\
1000 &$2.459(17)\times 10^{   5}$&$3.37(46)\times 10^{   2}$&$-8.3(11)\times 10^{   1}$&$2.53(47)\times 10^{   2}$\\ \\
\hline \\
  25 &$-19.602(61)$&$0.362(10)$&$-0.2949(86)$&$0.168(13)$\\
  50 &$-23.0686(88)$&$0.4622(69)$&$-0.2457(36)$&$0.2185(78)$\\
 100 &$-19.58(21)$&$0.551(11)$&$-0.1952(42)$&$0.366(12)$\\
 300 &$46.12(55)$&$0.657(20)$&$-0.1938(60)$&$0.454(21)$\\
 500 &$150.31(62)$&$0.651(39)$&$-0.172(10)$&$0.489(41)$\\
 700 &$274.2(14)$&$0.657(56)$&$-0.164(14)$&$0.482(57)$\\
1000 &$481.3(33)$&$0.669(90)$&$-0.163(22)$&$0.506(93)$\\
\end{tabular}
 \end{ruledtabular}
 \end{table*}
 \begin{table*}
 \caption{The same as Table~\ref{tabU2} but for the ${\rm Bi} \, {\rm Au}^{161+}$ quasimolecule. \label{tabBiAu}}
 \begin{ruledtabular}
 \begin{tabular}{lllll}
 Distance [fm]&
 \multicolumn{1}{c}{${\mathcal E}_a$}&
 \multicolumn{1}{c}{$\Delta E_{\rm SE}$}&
 \multicolumn{1}{c}{$\Delta E_{\rm VP}$} &
 \multicolumn{1}{c}{$\Delta E_{\rm QED}$} \\[0.1cm] \colrule
 \\[-0.15cm]
  15 &$-2.378(10)\times 10^{   5}$&$6.955(25)\times 10^{   3}$&$-6.545(23)\times 10^{   3}$&$4.10(34)\times 10^{   2}$\\
  25 &$-1.730(27)\times 10^{   5}$&$5.3618(71)\times 10^{   3}$&$-3.6905(49)\times 10^{   3}$&$1.6713(86)\times 10^{   3}$\\
  50 &$-7.990(27)\times 10^{   4}$&$3.549(20)\times 10^{   3}$&$-1.726(10)\times 10^{   3}$&$1.822(23)\times 10^{   3}$\\
 100 &$6.31(94)\times 10^{   3}$&$2.229(43)\times 10^{   3}$&$-8.91(17)\times 10^{   2}$&$1.337(46)\times 10^{   3}$\\
 300 &$1.376(10)\times 10^{   5}$&$9.21(28)\times 10^{   2}$&$-2.549(80)\times 10^{   2}$&$6.66(30)\times 10^{   2}$\\
 500 &$1.9836(13)\times 10^{   5}$&$5.61(17)\times 10^{   2}$&$-1.374(43)\times 10^{   2}$&$4.24(18)\times 10^{   2}$\\
 700 &$2.3746(22)\times 10^{   5}$&$3.96(18)\times 10^{   2}$&$-9.23(44)\times 10^{   1}$&$3.04(19)\times 10^{   2}$\\
1000 &$2.7695(48)\times 10^{   5}$&$2.75(23)\times 10^{   2}$&$-6.22(53)\times 10^{   1}$&$2.13(24)\times 10^{   2}$\\ \\
\hline \\
  15 &$-6.983(30)$&$0.20428(73)$&$-0.19213(69)$&$0.0120(10)$\\
  25 &$-8.47(13)$&$0.26232(34)$&$-0.18065(23)$&$0.08277(42)$\\
  50 &$-7.818(26)$&$0.3472(20)$&$-0.16893(98)$&$0.1783(22)$\\
 100 &$1.24(28)$&$0.4363(84)$&$-0.1755(33)$&$0.2628(90)$\\
 300 &$80.81(62)$&$0.541(16)$&$-0.1506(46)$&$0.391(17)$\\
 500 &$194.09(13)$&$0.559(17)$&$-0.1355(41)$&$0.425(17)$\\
 700 &$325.29(30)$&$0.543(25)$&$-0.1264(60)$&$0.426(26)$\\
1000 &$541.99(05)$&$0.549(46)$&$-0.121(10)$&$0.427(47)$\\
\end{tabular}
 \end{ruledtabular}
 \end{table*}
\end{widetext}

\section{Summary and outlook}
\label{summary}

In summary, we presented a theoretical approach for {\it ab initio} calculations of QED corrections to (quasi) molecular energy levels. In our approach, we generate molecular wave functions in spherical coordinates in terms of an expansion over the monopole solutions of Dirac equation. Based on such representation of wave functions, we employ the standard---for atomic physics---methods for the evaluation of the first--order self--energy and vacuum--polarization corrections. In order to illustrate the application of the proposed approach, we computed the Dirac energy and QED corrections for the 1$\sigma_g$ ground state of U$_2^{183+}$, ${\rm U} \, {\rm Pb}^{173+}$ and ${\rm Bi} \, {\rm Au}^{161+}$ dimers. Such quasimolecular systems can be produced in slow ion--ion collisions and are used as a tool for studying QED effects in the presence of (sub--) critical electromagnetic fields. Our calculations were performed for various inter--nuclear distances $R$, thus allowing us to explore QED corrections both in ``molecular'' and in a ``single ion'' regime. The relative accuracy of the calculations was attributed mainly to the truncation of the partial--wave expansion of molecular wave function, and was found not to exceed 10~\% even for the most problematic case of large inter--nuclear distances. Based on these findings we argue that even moderate basis set of partial wave functions can be used for accurate QED calculations of (quasi) molecular systems in spherical coordinates.

In the present work, we focused on the analysis of QED corrections for dimer quasimolecules. However, the developed approach can be extended to describe more complex molecules, composed of many atoms, and which usually do not posses axial symmetry. To explore such ``multi--center'' systems, one has to modify multipole expansions of both, the interaction potential and the molecular wave function, by adding summation over the angular momentum projections. Although making the calculations more demanding, it will open a promising route for the application of the developed approach for quantum chemistry purposes.



\end{document}